# Planck-Einstein-de Broglie relations for wave packet: the acoustic world

Ion Simaciu[1,a], Zoltan Borsos[1,b], Gheorghe Dumitrescu[2], Nan Georgeta[1]

[1] Petroleum-Gas University of Ploieşti, Ploieşti 100680, Romania

[2] High School Toma N. Socolescu, Ploieşti, Romania

## Abstract

*In this paper we study the relations of Planck-Einstein-de Broglie type for the wave packets. We assume that the wave packet is a possible model of particle. When studying the behaviour of the wave packet for standing waves, in relation to an accelerated observer (i.e. Rindler observer), there can be demonstrated that the equivalent mass of the packet is the inertial mass. In our scenario, the waves and of the wave packets are depicted by the strain induced/produced in the medium. The properties of the waves, of the wave packet and, generally, of the perturbations in a material medium suggest the existence of an acoustic world. The acoustic world has mechanical and thermodynamical properties. The perturbations that are generated and propagated in the medium are correlated by means of acoustic waves with maximum speed. The observers of this world of disturbances (namely the acoustic world) have senses that are based on the perception of mechanical waves (disturbance of any kind) and apparatus for detecting and acquiring information by means of the same type of wave. Time and length measurements (and other parameters) are correlated by Lorentz type transformations, where the maximum speed is the speed of the sound waves. By applying these transformations to the packet of standing waves, there results that the energy, mass, linear momentum and the action variable/variable action undergo relativistic changes. We highlight the fact that the dynamic relativist relationship between energy and momentum is a consequence of the wave packet model for a particle. We have also emphasized the existence of certain limits for the energy of the disturbance and the corresponding action variable.*

Keyword: wave packet, Rindler observer, acoustic world, Planck-Einstein-de Broglie relation

PACS: 45.20.-d , 47.10.-g , 62.30.+d, 03.30.+p

## Introduction

The aim of this paper is to infer the Planck-Einstein-de Broglie type relations [1] for mechanical wave packet. Also we study the properties of the acoustic world. The acoustic world is composed of disturbances that propagates in a fluid or solid: waves, wave packets, cavities etc.

The hypothesis of E. Schrödinger [2] that the particle can be modelled as a packet of standing waves in the configuration space has been reiterated by several authors. In these models, the waves which compose the packet are physical waves (electromagnetic or of other nature). Related to a moving reference, it becomes a packet modulated by a phase wave with the wave vector equal to the de Broglie wave vector $\vec{k}_B = \vec{p}/\hbar$ [3, 4, 5, 6]. The hypothesis that the de Broglie wave has an electromagnetic nature is also stated by M. Surdin in the field of stochastic electrodynamics [7]. In this stochastic model, the particle absorbs and emits waves from CZPF, waves that compose the phase wave on their direction of motion. The model of the standing waves packet is compatible with the cavity-guide model if the tridimensional packet consists of captured standing waves (cavity modes) [8, 9]. Louis de Broglie, in order to explain quantum effects as an interaction of microscopic systems with a sub-quantum environment, proposed a thermodynamics of the isolated particle (i.e. the hidden thermodynamics of isolated particles) in which the particle is modelled as a thermodynamic system in interaction with the sub-quantum medium (the stochastic medium) [10].

[a] sion@upg-ploiesti.ro; isimaciu@yahoo.com

[b] borsos.zoltan@gmail.com

In the first section, the authors study the wave packet as a macroscopic oscillator having a size equal to the average wavelength. We identify Einstein type relations between the energy and the equivalent mass of the packet. There is also a relation between the angular frequency and energy similar to the Planck relationship. Related to a moving observer, the packet acquires a linear momentum, as the product of the equivalent mass and the speed of the observer. Between the linear momentum and the action variable there is a de Broglie type of relation. Disturbances that are generated and propagate in the medium correlate by means of the waves that move at maximum speed. The observers from the world of such disturbances (i.e. the acoustic world) have senses that are based on the perception of the mechanical waves (namely disturbances of any kind) and detectors working by means of the same kind of waves [11]. Time and length measurements (and other parameters) are conected by Lorentz type transformations, where the maximum speed is the speed of the sound waves. When applying these transformations to the packet of standing waves, there results that energy, mass, linear momentum and *action variable* or the canonical action variable or the action-angle variable undergo relativistic changes.

In the second section, we study the behaviour of the standing wave packet related to an accelerated observer (i.e. Rindler observer). Through this process, we prove that *the equivalent mass of the wave packet is the inertial mass*.

In the third section, we study the dependency of the waves and of the wave packet parameters on the maximum deformation produced by them in the medium in which propagates through. It is pointed out that the following parameters: energy, equivalent mass and momentums are proportional to the square of the maximum deformation produced by these disturbances in the environment. The deformation in the medium is dependent upon the point.

In the section 4, we calculate local parameters corresponding to the disturbances of the environment: energy density, equivalent mass density, the density of the linear momentum and action variable.

In the fifth section, we analyze the thermodynamic properties of acoustic world. We highlight the fact that the dynamic, relativistic-type relationship between the energy and the linear momentum results from the wave packet model for a particle. We also reveal the existence of some limits for the disturbance energy and its corresponding action.

Section 6 includes the conclusions of the paper.

# 1.  Particle as a packet of standing waves

The assumption that a particle is a wave packet is a hypothesis that was formulated in the works of Schrödinger, according to whom the waves forming the packet were the de Broglie type waves [2, 12 Ch. 2.4]. Researchers Bernhard Haisch and Alfonso Rueda [6], when trying to model mechanical inertia as interaction with CZPF, also refer to the hypothesis of modelling the particle as a packet of standing waves and cites Kracklauer, Al. F., Luis de la Pena and Ana Maria Cetto [4, 5, 13]. In these works, the waves forming the packet were real waves (for example, electromagnetic). Stimulated by this king of research and considering that they confirm our unpublished works, cited in [8], we have decided to deepen this model [14, Ch. 2.5]. It is surprising that, in the works of Claude Elbaz [3, 15] prior to those cited by Haisch, it were included features and important developments of this model. The most important is the connection between the electromagnetic and gravitational interactions [16], by means of the wave packet model of standing waves.

In the acoustic world of an medium through which there propagate mechanical waves, the disturbances of the medium interact. Some disturbances can be modelled as acoustic cavities for the mechanical waves (non-homogenous and non-isotropic areas where the acoustic refraction index depends on the point). In liquid or solid media, there are cavities produced as a result of the connections breaking or, of the liquid transforming to fluid and of the solid turning liquid. To support such statements, we mention the papers that analyze, both theoretically and experimentally, the acoustic radiation forces or Bjerknes forces [17]. The interaction between a bubble (cavity in the liquid) and a sound wave is named the primary Bjerknes force [18]. The interaction between two pulsating bubbles is named the secondary Bjerknes [18]. There were also studied the forces induced by the field of sound waves. The *forces are induced by a sound wave field on dispersed particles (liquid drops, particles of dust, etc.) suspended in a fluid* [19].

In this section, we shall apply the results presented in the paper [1] to the packet of standing mechanical waves. This application is suggested and supported by the works of William George Unruh, Theodore Jacobson, Matt Visser, Harold E. Puthoff and those who follow them [20, 21, 22, 23, 24, 11], in which the fluctuations of an elastic environment, considered events, are correlated by means of Lorentz type transformations, where the maximum speed is the speed of the mechanical waves, $c \to u$, $\beta = \upsilon/u$.

Let's consider a packet of standing waves with the formula

$$\Psi_0(z_0, t_0) = A_0(k_0 z_0) \psi(\omega_0 t_0) \tag{1}$$

where

$$(\Delta \omega)_0 = (\omega_{av})_0 = \omega_0 \text{ și } (\Delta k)_0 = (k_{av})_0 = k_0 \tag{2}$$

so that the spatial extension (size $(\Delta z)_0$ where the amplitude of oscillations does not equal zero), related to the reference frame against which the packet is at rest

$$(\Delta k)_0 (\Delta z)_0 = k_0 (\Delta z)_0 = 2\pi \text{ sau } (\Delta z)_0 = \frac{2\pi}{k_0} = \lambda_0, \tag{3}$$

With $\lambda_0$ the average wavelength. When related to a moving observer, with the speed $\upsilon$, the standing packet becomes a moving packet [15]

$$\Psi(z,t) = A(kz - \beta \omega t) \psi(\omega t - \beta k z) \tag{4}$$

with the group velocity $\upsilon_g = \upsilon = \beta u$, $\beta = \upsilon/u$, the phase speed $\upsilon_\varphi = u^2/\upsilon = u/\beta$ and the spatial

extension

$$(\Delta k)(\Delta z) = k(\Delta z) = 2\pi \quad \text{sau} \quad \Delta z = \frac{2\pi}{k} = \lambda = \lambda_0\sqrt{1-\beta^2} < \lambda_0. \tag{5}$$

The moving packet is characterized by wave averaged parameters and modulated parameters:

$$\omega_{av} = \frac{\omega_0}{\sqrt{1-\beta^2}} = \gamma\omega_0 = \omega \quad \text{and} \quad k_{av} = \beta k = \gamma\beta k_0 = k_B, \tag{6}$$

$$\omega_{mod} = \Delta\omega = \gamma\beta\omega_0 = \beta\omega \quad \text{and} \quad k_{mod} = \Delta k = \gamma k_0 = k. \tag{7}$$

According to the results from the previous sections, each element with the length $dz$ and mass $dm = \rho l_x l_y dz$ performs an oscillating motion with the angular frequency $\omega_0$, the amplitude

$$q_p(z_0, t_0) = q_{0p} A_0(k_0 z_0) \psi(\omega_0 t_0) \tag{8}$$

and the oscillation velocity

$$\dot{q}_p(z_0, t_0) = q_{0p} A_0(k_0 z_0) \dot{\psi}(\omega_0 t_0) = q_{0p} \omega_0 A_0(k_0 z_0) \frac{\partial \psi(\omega_0 t_0)}{\partial(\omega_0 t_0)} \tag{9}$$

Mass element $dm$ has the kinetic energy

$$dE_k = \frac{1}{2} dm q_{0p}^2 \omega_0^2 A_0^2(k_0 z_0) \left[\frac{\partial \psi(\omega_0 t_0)}{\partial(\omega_0 t_0)}\right]^2 \tag{10}$$

and the potential energy

$$dE_p = \frac{1}{2} dm q_{0p}^2 \omega_0^2 A_0^2(k_0 z_0) \psi^2(\omega_0 t_0). \tag{11}$$

Therefore, the total energy of the mass element is

$$dE = \frac{1}{2} dm q_{0p}^2 \omega_0^2 A_0^2(k_0 z_0) \left\{\psi^2(\omega_0 t_0) + \left[\frac{\partial \psi(\omega_0 t_0)}{\partial(\omega_0 t_0)}\right]^2\right\}. \tag{12}$$

The average, over time, value of this energy is

$$\langle dE \rangle_t = \frac{1}{2} dm q_{0p}^2 \omega_0^2 A_0^2(k_0 z_0) \frac{1}{T_0} \int_0^{T_0} \left\{\psi^2(\omega_0 t_0) + \left[\frac{\partial \psi(\omega_0 t_0)}{\partial(\omega_0 t_0)}\right]^2\right\} dt_0. \tag{13}$$

The total energy of the packet is

$$E_{p\lambda_0} = \int_{-\lambda_0/2}^{\lambda_0/2} \langle dE \rangle_{tz} = \frac{1}{2} \rho l_x l_y q_{0p}^2 \omega_0^2 \left[\int_{-\lambda_0/2}^{\lambda_0/2} A_0^2(k_0 z_0) dz_0\right] \frac{1}{T_0} \int_0^{T_0} \left\{\psi^2(\omega_0 t_0) + \left[\frac{\partial \psi(\omega_0 t_0)}{\partial(\omega_0 t_0)}\right]^2\right\} dt_0. \tag{14}$$

The integral

$$\int_{-\lambda_0/2}^{\lambda_0/2} A_0^2(k_0 z_0) dz_0 = \frac{(\Delta z)_0}{\alpha} = \frac{\lambda_0}{\alpha} \tag{15}$$

is proportional to the extension of the packet $(\Delta z)_0$ given by the relation (3). The proportionality constant $\alpha$ depends on the shape of the packet.

For a sine wave dependency over time, the integer in time from (14) is approximately equal to $T_0$. Therefore, the expression of the total energy of the packet becomes

$$E_{p\lambda_0} = \frac{1}{2\alpha} \rho l_x l_y q_{0p}^2 \omega_0^2 \lambda_0 . \tag{16}$$

If we express total energy, based on the propagation speed of the mechanical waves $u$, there results

$$E_{p\lambda_0} = \frac{2\pi^2 \rho l_x l_y q_{0p}^2}{\alpha \lambda_0} u^2 . \tag{17}$$

The expression (17) of the energy suggests the existence of a mechanical inertia of the wave packet that can be measured by the equivalent rest mass of the packet

$$m_{p\lambda_0} \equiv \frac{E_{p\lambda_0}}{u^2} = \frac{2\pi^2 \rho l_x l_y q_{0p}^2}{\alpha \lambda_0} = \frac{m_{\lambda_0}}{2\alpha} \left( \frac{2\pi q_{0p}}{\lambda_0} \right)^2 \tag{18}$$

with $m_{\lambda_0} = \rho \lambda_0 l_x l_y$ being the rest mass of the substance corresponding to the packet. We notice that to the same energy, there correspond different mechanical inertia ( $E_{p\lambda_0} = (m_{\lambda_0}/2) \langle \dot{q}_{0p}^2 \rangle_{tz} = m_{p\lambda_0} u^2$ ).

To this mass there also corresponds an internal linear momentum of the waves belonging the packet, waves that propagate in opposite directions with the speed $u$

$$P_{p0} = m_{p\lambda_0} u = \frac{2\pi^2 \rho l_x l_y q_{0p}^2 u}{\alpha \lambda_0} = \left( \frac{2\pi q_{0p}}{\lambda_0} \right)^2 \frac{m_{\lambda_0}}{2\alpha} u \tag{19a}$$

which is different from the square momentum of the microscopic oscillators corresponding to the packet

$$P_{p\lambda_0} = N_{p\lambda_0} P_a = N_{p\lambda_0} \sqrt{\langle p_{ap}^2 \rangle_{z,t}} = N_{p\lambda_0} \frac{1}{\sqrt{2\alpha}} m_a q_{0p} \omega_0 = \frac{\sqrt{2}\pi \rho l_x l_y q_{0p} u}{\sqrt{\alpha}} = 2\sqrt{2\alpha} \left( \frac{\lambda_0}{2\pi q_{0p}} \right) P_{p0}, \tag{19b}$$

with $\alpha$ the constant that occurs at spacial averaging and depends on the shape of the packet.

If we consider the standing packet as a macroscopic oscillator, then, to its energy, given by the relation (13), there corresponds an action variable, which, according to Hamilton's formalism has cyclic coordinates [25, p. 436],

$$J_{p0} = \frac{E_{p\lambda_0}}{\omega_0} = \frac{1}{2\alpha} \rho l_x l_y q_{0p}^2 \lambda_0 \omega_0 = \frac{\pi}{\alpha} \rho l_x l_y q_{0p}^2 u . \tag{20}$$

The relations (20) for packet as macroscopic oscillator is similar to the Planck's relation [12, Ch. 1.1; 30, Ch. 41.3].

With respect to an observer moving at the speed $\upsilon$, the packet has the energy

$$E_{p\lambda} = \frac{1}{2\alpha} \rho l_x l_y q_{0p}^2 \omega^2 \lambda = \frac{\pi^2}{\alpha} (\rho l_x l_y \lambda) \frac{q_{0p}^2}{\lambda^2} u^2 = \gamma E_{p\lambda_0} , \tag{21}$$

the equivalent relativistic mass

$$m_{p\lambda} = \frac{2}{\alpha}\pi^2\left(\rho l_x l_y \lambda\right)\frac{q_{0p}^2}{\lambda^2} = \gamma m_{p\lambda_0} > m_{p\lambda_0},\tag{22}$$

the internal linear momentum

$$\begin{aligned}\left\langle \vec{P}_p^2 \right\rangle_{tz} &= \left\langle \left(\vec{P}_{pT} + \vec{P}_{po} + \vec{p}_{pk}\right)^2 \right\rangle_{tz} = \\ &\left(\rho l_x l_y \lambda\right)^2 \left[\left\langle \dot{\vec{q}}_T^2 \right\rangle_t + \left\langle \dot{\vec{q}}_p^2 \right\rangle_{tz} + \left\langle \vec{\upsilon}^2 \right\rangle_{tz} + 2\left(\left\langle \dot{\vec{q}}_T \dot{\vec{q}}_p \right\rangle_{tz} + \left\langle \dot{\vec{q}}_T \vec{\upsilon} \right\rangle_{tz} + \left\langle \dot{\vec{q}}_p \vec{\upsilon} \right\rangle_{tz}\right)\right] = \\ &\left(\rho l_x l_y \lambda\right)^2 \left[\left\langle \dot{\vec{q}}_T^2 \right\rangle_{tz} + \left\langle \dot{\vec{q}}_p^2 \right\rangle_{tz} + \vec{\upsilon}^2\right] = \left(\rho l_x l_y \lambda_0\right)^2 \frac{\vec{q}_{0T}^2 \omega_a^2}{2} + P_{p\lambda}^2 + m_{p\lambda}^2 \vec{\upsilon}^2\end{aligned}\tag{23}$$

the square momentum of the microscopic oscillators corresponding to the packet

$$P_{p\lambda} = \gamma P_{p\lambda_0}\tag{24}$$

and *the action variable*

$$J_p = \frac{E_{p\lambda}}{\omega} = \frac{1}{2}\rho l_x l_y q_{0p}^2 \lambda\omega = \pi\rho l_x l_y q_{0p}^2 u = J_{p0}.\tag{25}$$

We can also express the action variable of the packet through the atomic action variable (eq. 33, from paper [1]), with $\omega_j \to \omega$

$$J_p = \frac{E_{p\lambda}}{\omega} = \frac{1}{2}\rho l_x l_y q_{0p}^2 \lambda\omega = \pi\rho l_x l_y q_{0p}^2 u = J_{p0} = \left(n l_x l_y \lambda\right)\left(\frac{\pi}{\lambda}m_a q_{0p}^2 u\right) = N_{packet}J_a.\tag{26}$$

From eq. (25) it follows that *the action variable is not only adiabatic invariant* [16] but also *Lorentz invariant* [26].

With respect to the moving observer, the packet transports an energy which is the kinetic energy of the packet and it has a linear momentum

$$p_p = m_{p\lambda}\upsilon = \frac{2\pi^2 \rho l_x l_y q_{0p}^2 \upsilon}{\alpha\lambda} = \left(\frac{2\pi q_{0p}}{\lambda_0}\right)^2 \frac{m_\lambda}{2\alpha}\upsilon.\tag{27}$$

In a first approximation, the kinetic energy is expressed

$$E_{k\lambda} = \frac{p^2}{2m_{p\lambda}} = \frac{m_{p\lambda}\upsilon^2}{2}.\tag{28}$$

This energy can be put in relation with the energy of the packet (of the microscopic oscillators in the volume $\Delta z l_x l_y = \lambda l_x l_y$) which oscillates with the modulation angular frequency $\omega_{\text{mod}} = \beta\omega$

$$E_\lambda(\omega_{\text{mod}}) = \frac{1}{2\alpha}\rho l_x l_y q_0^2 \omega_{\text{mod}}^2 \lambda = \frac{1}{2\alpha}\rho l_x l_y q_{0p}^2 \beta^2 \omega^2 \lambda = \frac{\rho l_x l_y \lambda q_{0p}^2 \omega^2}{2\alpha u^2}\upsilon^2 = m_\lambda \upsilon^2,\tag{29}$$

Comparing (29) and (28), it follows

$$E_\lambda(\omega_{\text{mod}}) = 2E_{c\lambda}.\tag{30}$$

Just like in the theory of relativity, the kinetic energy of the packet is the difference between its relativistic energy and its rest energy

$$E_{c\lambda} = E_\lambda(\omega) - E_{\lambda_0}(\omega_0) = \left[m_\lambda(\omega) - m_{\lambda_0}(\omega_0)\right] u_l^2. \tag{31}$$

The ratio of the action variable (25) and the linear momentum of the packet (27) is

$$\frac{J}{p} = \frac{J_0}{p} = \frac{\dfrac{1}{2\alpha}\rho l_x l_y q_{0p}^2 \lambda \omega}{\dfrac{2\pi^2 \rho l_x l_y q_{0p}^2 \upsilon}{\alpha \lambda}} = \frac{u_l^2}{\upsilon \omega} = \frac{1}{k_{av}} = \frac{\lambda_{av}}{2\pi}, \tag{32}$$

that is a *de Broglie type relation*, if

$$J 2\pi = J_0 2\pi \tag{33}$$

is a *Planck type constant for action*, corresponding to the elastic medium.

## 2. The inertial mass of the wave and of the wave packet

To derive the physical significance of the wave mass given by the equation (42) from paper [1], we will derive the expression of the wave energy and mass with respect to an accelerated frame of reference.

Let $S'(x', y', z', t')$ be an accelerated frame of reference characterized by the acceleration vector $\vec{a}(0, 0, a_z = a)$ with respect to an inertial frame of reference $S(x, y, z, t)$. The relations between $z$ and $t$ and the proper time, $\tau$, in the instantaneous rest frame of the accelerated observer (Rindler observer, [27, 28]) are

$$t(\tau) = \frac{u}{a}\sinh\left(\frac{a\tau}{u}\right), \quad z(\tau) = \frac{u^2}{a}\cosh\left(\frac{a\tau}{u}\right), \tag{34}$$

In the instantaneous rest frame of the observer, the angular frequency $\omega'$ has the equation

$$\omega'(\tau) = \frac{\omega - k\upsilon(\tau)}{\sqrt{1 - \upsilon^2(\tau)/u^2}} = \omega \exp\left(\frac{-a\tau}{u}\right), \quad k = \frac{\omega}{u}, \tag{35a}$$

for the waves which propagate in the observer's acceleration direction and

$$\omega'(\tau) = \omega \exp\left(\frac{a\tau}{u}\right), \quad k = \frac{-\omega}{u}, \tag{35b}$$

for the waves which propagate in the opposite direction of the observer's acceleration.

To derive the equation of the wave energy corresponding to a half-wavelength (equation 36 from paper [1]) with respect to the Rindler observer we take the action variable to be, according to equation (25), Lorentz-invariant

$$J'_{\lambda/2} = J_{\lambda/2}. \tag{36}$$

Under these conditions, the energy equation (eq. 36 from paper [1]) transforms in the same way as the angular frequency (35)

$$E'\lambda/2 = J'_{\lambda/2}\omega' = J_{\lambda/2}\omega'. \tag{37}$$

Substituting (35) in equation (37), results

$$E'_{\lambda/2} = J_{\lambda/2}\omega \exp\left(\frac{a\tau}{u}\right) = E_{\lambda/2}\exp\left(\frac{a\tau}{u}\right). \tag{38}$$

The corresponding power with respect to the Rindler observer is

$$P'_{\lambda/2} = \frac{dE'_{\lambda/2}}{d\tau} = E_{\lambda/2}\left(\frac{\mp a}{u}\right)\exp\left(\frac{\mp a\tau}{u}\right) = F'_{\lambda/2}u. \tag{39}$$

From the equation (39), it follows

$$F'_{\lambda/2} = \frac{E_{\lambda/2}}{u^2}(\mp a)\exp\left(\frac{\mp a\tau}{u}\right). \tag{40}$$

Substituting the expression of the equivalent mass, according to the relation (41) from paper [1], in the equation (40), it follows

$$F'_{\lambda/2} = (\mp a)m_{u,\lambda/2}\exp\left(\frac{\mp a\tau}{u}\right) = (\mp a)m'_{u,\lambda/2}. \tag{41}$$

From the equation (41), it follows that the inertial mass with respect to the Rindler observer is dependent on the proper time $\tau$

$$m'_{u,\lambda/2}(a,\tau) = m_{u,\lambda/2}\exp\left(\frac{\mp a\tau}{u}\right). \tag{42}$$

and $m_{u,\lambda/2} = m'_{u,\lambda/2}(a=0)$, therefor the wave inertial mass with respect to the unaccelerated observer $S(x,y,z,t)$, is the equivalent mass of the wave. One can notice that the wave mass transforms in the same way as the energy and the angular frequency.

The mass of the wave packet has the same properties, that is, *the equivalent mass of the packet is the inertial mass of the packet.*

## 3. The dependence of the waves and of the wave packet parameters on the strain

The specific parameters of the waves and of the wave packet can be expressed depending on the fractional stretch or the strain caused by them in the medium through propagate.

By definition [29, Ch. 3.8.1], the fractional stretch or the strain is

$$\varepsilon(z,t) = \frac{\partial q_z}{\partial z} = \frac{\partial q_z}{\partial t}\frac{1}{(\partial z/\partial t)} = -\frac{\dot{q}_z}{u} \tag{43}$$

and, consequently, can be expressed as the ratio between the oscillation velocity of the medium particles and the propagation velocity of the strains in that medium. For the longitudinal waves the strain is negative in the compression areas ( $\varepsilon < 0$, the direction of the oscillation velocities is the same as that of the propagation velocity) and positive in the decompression areas ( $\varepsilon > 0$, the direction of the oscillation velocities is the opposite of that of the propagation velocity). For the transverse waves the strain has the same sign.

In the case of the propagating wave, $\Psi(z,t) = q_z = q_{0z}\sin(\omega t - kz)$, the strain is

$$\varepsilon(z,t) = \frac{\partial q_z}{\partial z} = -q_{0z}k\cos(\omega t - kz) = -\frac{2\pi q_{0z}}{\lambda}\cos(\omega t - kz), \qquad (44)$$

with $\varepsilon_0 = q_{0z}k = 2\pi q_{0z}/\lambda$ maximum strain. The average strain in time and space is null and the averaged square of the strain in time and space is

$$\left\langle \varepsilon^2 \right\rangle_{zt} = \frac{1}{2}\left(\frac{2\pi q_{0z}}{\lambda}\right)^2 = \frac{1}{2}\varepsilon_0^2. \qquad (45)$$

For the case of the standing wave, $\Psi_s(z,t) = q_{sz} = q_{0sz}\sin(kz)\sin(\omega t)$, the strain is

$$\varepsilon_s(z,t) = \frac{\partial q_z}{\partial z} = kq_{0sz}\sin(kz)\sin(\omega t) = \frac{2\pi q_{0sz}}{\lambda}\cos(kz)\sin(\omega t), \qquad (46)$$

with $\varepsilon_{0s} = q_{0sz}k = 2\pi q_{0sz}/\lambda$ maximum strain. The average strain in time and space is null and the averaged square of the strain in time and space is

$$\left\langle \varepsilon_s^2 \right\rangle_{szt} = \frac{1}{4}\left(\frac{2\pi q_{0sz}}{\lambda}\right)^2 = \frac{1}{4}\varepsilon_{0s}^2. \qquad (47)$$

In the case of standing packet given by the relation (8), with a sinusoidal dependence in time, the strain is

$$\varepsilon_{0p}(z,t) = k_0 q_{0p} \frac{\partial A_0(k_0 z_0)}{\partial(k_0 z_0)}\sin(\omega t) = \frac{2\pi q_{0p}}{\lambda_0}\frac{\partial A_0(k_0 z_0)}{\partial(k_0 z_0)}\sin(\omega t) \qquad (48)$$

with $\varepsilon_{0p} = q_{0p}k = 2\pi q_{0p}/\lambda$ maximum strain. The average strain in time and space is null and the averaged square of the strain in time and space is

$$\left\langle \varepsilon_p^2 \right\rangle_{z,t} = \frac{1}{2\alpha}\left(\frac{2\pi q_{0p}}{\lambda}\right)^2 = \frac{1}{2\alpha}\varepsilon_{0p}^2. \qquad (49)$$

With these results we obtain the following dependances: the wave energy given by the relation (40) from paper [1] becomes

$$E_{\lambda/2} = \frac{\varepsilon_0^2}{2}\left(m_{\lambda/2}u^2\right) = \left\langle \varepsilon_0^2 \right\rangle_{z,t}\left(m_{\lambda/2}u^2\right); \qquad (50)$$

the equivalent mass of the macroscopic oscillator, corresponding to a half-wavelength given by eq. (41) from paper [1], becomes

$$m_{u,\lambda/2} = \frac{\varepsilon_0^2}{2}m_{\lambda/2} = \left\langle \varepsilon_0^2 \right\rangle_{z,t} m_{\lambda/2} \qquad (51)$$

and the *action variable* becomes

$$J_{\lambda/2} = \frac{\lambda}{2\pi}\left(\left\langle \varepsilon_0^2 \right\rangle_{z,t} m_{\lambda/2}\right)u = \frac{\lambda}{2\pi}m_{u,\lambda/2}u \qquad (52a)$$

or

$$2\pi J_{\lambda/2} = \lambda\left(\left\langle \varepsilon_0^2 \right\rangle_{z,t} m_{\lambda/2}\right)u_l = \lambda m_{u,\lambda/2}u, \qquad (52b)$$

similar to the equation for the *Compton wavelength* [12, Ch. 1.3].

In the case of the rest wave packet we obtain the following expressions: the packet energy

$$E_{p\lambda_0} = \frac{2\pi^2 \rho l_x l_y q_{0p}^2}{\alpha \lambda_0} u^2 = \langle \varepsilon_p^2 \rangle_{z,t} m_{\lambda_0} u^2 = m_{p0} u^2 \tag{53}$$

the equivalent mass of the packet

$$m_{p0} = \frac{\varepsilon_{0p}^2}{2\alpha} m_{\lambda_0} = \langle \varepsilon_p^2 \rangle_{z,t} m_{\lambda_0} \tag{54}$$

and the action variable of the packet

$$J_{p0} = \frac{\lambda_0}{2\pi} \left( \langle \varepsilon_p^2 \rangle_{z,t} m_{\lambda_0} \right) u = \frac{\lambda_0}{2\pi} m_{p0} u \tag{55a}$$

or

$$2\pi J_{p0} = \lambda_0 m_{p0} u . \tag{55b}$$

In the case of a wave packet related to a moving observer, the maximum strain is, at least for the longitudinal waves, a Lorentz invariant

$$\varepsilon_{0p} = \frac{2\pi q_{0p}}{\lambda_0} = \frac{2\pi q_{0p\upsilon}}{\lambda} = \varepsilon_p . \tag{56}$$

With respect to the moving observer we have the following expressions: the packet energy

$$E_{p\lambda} = \frac{2\pi^2 \rho l_x l_y q_{0p}^2}{\alpha \lambda} u^2 = \langle \varepsilon_p^2 \rangle_{z,t} m_\lambda u^2 = m_p u^2 ; \tag{57}$$

the equivalent mass of the packet

$$m_p = \frac{\varepsilon_p^2}{2\alpha} m_\lambda = \langle \varepsilon_p^2 \rangle_{z,t} m_\lambda \tag{58}$$

and the *action variable* of the packet

$$J_p = \frac{\lambda}{2\pi} \left( \langle \varepsilon_p^2 \rangle_{z,t} m_\lambda \right) u = \frac{\lambda}{2\pi} m_p u \tag{59a}$$

or

$$2\pi J_p = \lambda m_p u . \tag{59b}$$

One can notice that all these quantities are proportional with the square of the maximum strain $\varepsilon^2$ caused by the waves in the propagation medium.

## 4. Specific parameters

We can derive the wave and packet local parameters, independent of the cross dimensions.

Dividing relations (50-52) by the corresponding volume, we obtain the following specific parameters: the average energy density in the wave

$$w_{u\lambda/2} \equiv \frac{E_{\lambda/2}}{l_x l_y (\lambda/2)} = (\rho u^2) \frac{\varepsilon_0^2}{2} = (\rho u^2) \langle \varepsilon_0^2 \rangle_{z,t} ; \tag{60}$$

*equivalent mass density* of the macroscopic corresponding to a half-wavelength given by eq. (41) from paper [1] becomes

$$\rho_{u,\lambda/2} = \rho \frac{\varepsilon_0^2}{2} = \rho \langle \varepsilon_0^2 \rangle_{z,t} \tag{61}$$

and the action variable density

$$w_{J_{\lambda/2}} \equiv \frac{J_{\lambda/2}}{l_x l_y (\lambda/2)} = \frac{\lambda}{2\pi} \left( \rho \langle \varepsilon_0^2 \rangle_{z,t} \right) u = \frac{\lambda}{2\pi} \rho_{u,\lambda/2} u . \tag{62}$$

In the case of the rest wave packet, we obtain the following expressions: energy density of the packet

$$w_{p\lambda_0} \equiv \rho u^2 \left( \frac{\varepsilon_{0p}^2}{2\alpha} \right) = \rho u^2 \langle \varepsilon_p^2 \rangle_{z,t} ; \tag{63}$$

equivalent mass density of the packet

$$\rho_{p\lambda_0} = \rho \left( \frac{\varepsilon_{0p}^2}{2\alpha} \right) = \rho \langle \varepsilon_p^2 \rangle_{z,t} \tag{64}$$

and the action variable density of the packet

$$w_{J_{p0}} = \frac{\lambda_0}{\pi} \rho_{\lambda_0} \langle \varepsilon_p^2 \rangle_{z,t} u = \frac{\lambda_0}{2\pi} \rho_{p\lambda_0} u . \tag{65}$$

The linear momentum density, for the moving packet, is

$$w_{pp} = \rho_{p\lambda} \upsilon = \left( \frac{\varepsilon_p^2}{2\alpha} \right) \rho \upsilon = \langle \varepsilon_p^2 \rangle_{z,t} \rho \upsilon . \tag{66}$$

One can notice that the specific parameters are also proportional with the square of the maximum strain $\varepsilon_0^2$ or with the average of the square of the maximum strain $\langle \varepsilon^2 \rangle$.

## 5. The acoustic world

### 5.1. Thermodynamical properties of the acoustic world

Consider a homogeneous and isotropic medium (fluid or solid) with a mass density $\rho$, placed in an enclosure/ with a volume $V$ and with a temperature $T$. At the microscopic level, the medium is composed by identical particles with mass $m_a$, having the particle density $n = \rho / m_a$. In the case of the fluid, each particle has an average thermal energy [30, Ch. 39.5]

$$E_a = \frac{m_a \upsilon_T^2}{2} = \frac{3k_B T}{2} . \tag{67}$$

In the case of the solid, each particle executes a random (in sense and direction) oscillatory motion with the angular frequency given by eq. (21) from paper [1]. The average thermal energy of the microscopic oscillators is

$$E_{ao} = \frac{m_a q_T^2 \omega_a^2}{2} = k_B T \tag{68}$$

and the average thermal action variable is

$$J_{aT} = \frac{1}{2} m_a q_T^2 \omega_a , \qquad (69)$$

The perturbations (the waves and waves packets formed through interference–diffraction) in this medium constitute an „acoustic world" which correlates itself through the maximum velocity, which is the waves velocity in the medium, $u$. The oscillations in the material, caused by the perturbations, overlap the random thermal oscillations or the thermal motion. In this world, the group velocity of perturbations (the speed at which the amplitudes propagate) is less (or most equal) than the waves velocity, $\upsilon = \upsilon_g \leq u$.

We assume that localized identical perturbations (wave packets) in random motion are formed in the container. The mass of these perturbations is the equivalent mass defined by (18) and (22). These identical perturbations in random motion constitute an "acoustical" gas. If the particle density of the acoustical perturbations is $n_p$ and their temperature is $T_p$, then the pressure of the acoustical gas is [30, Ch. 39.5]

$$p_p = n_p k_B T_p , \qquad (70)$$

the average kinetic energy is

$$E_p = \frac{m_p \upsilon_{T_p}^2}{2} = \frac{3 k_B T_p}{2} . \qquad (71)$$

This thermal energy corresponds to the motion energy of a wave packet given by relation (28). The temperature of the „acoustical gas" is different from the temperature of the medium $T_p \neq T$. One can establish a connection between these temperatures on the basis of the conservation laws applied both to „acoustical" gas and to the medium in the enclosure with constant volume. In the relations (70) and (71) we used the Boltzmann constant $k_B$ because this constant, which appears in the expression of the average energy of the degree of freedom, doesn't depend on the worlds' level of structuration (in our case, the world of the medium and the „acoustic" world). The Boltzmann constant is the entropy quanta [31], that is the minimum entropy that the system can exchange. The nature/property of being the minimum quantity of entropy can be shown most simply in the case of a system with constant volume and temperature. The entropy variation can be achieved only through the substance exchange which can change the total number of degrees of freedom $Ni$

$$dS = \frac{dU}{T} = \frac{d\left(Ni \frac{k_B T}{2}\right)}{T} = \frac{k_B}{2} d(Ni) = \frac{k_B}{2} (i dN + N di) . \qquad (72)$$

In this case, the variation of the number of particles and of the number of degrees of freedom for each particle can be made only discontinuously. We notice that in open thermodinamical system the exchange of a particle (that have the same number of degrees of freedom, $\Delta i = 0$), implies a change of the particle number, $\Delta N = \pm 1$, and of the entropy with

$$\Delta S = \frac{k_B}{2} (i \Delta N) = \pm i \frac{k_B}{2} . \qquad (73)$$

The variation of the number of degrees of freedom is possible for a closed system, $\Delta N = 0$, with a fixed volume and constant temperature, if the temperature for the excitation of rotations and vibrations is reached

$$\Delta S = \frac{k_B}{2} N (\Delta i). \tag{74}$$

with the variation of the number of degrees of freedom an integer number (for instance, $\Delta i = 2$, if the rotation degrees for a diatomic molecule are added).

In the acoustical gas waves can be formed to propagate at a velocity $u_p$ different from the waves velocity in the gaseous medium, $u_a$, which has the temperature $T$

$$u_p = \sqrt{\frac{\gamma k_B T_p}{m_p}} < \sqrt{\frac{\gamma k_B T}{m_a}} = u_a. \tag{75}$$

Restricting to the acoustic gas frame one can set for the acoustic gas a thermodynamic theory which models the chaotic behaviour of the wave packets (atoms/molecules of the acoustical gas). For us, as observers, there is also a temperature corresponding to the medium which makes possible the existence of the wave packets. Hence we can understand/interpret a thermodynamic of the isolated particle [10].

We presented this gaming of the worlds/with the worlds in order to show the fact that one can structure worlds in which the parameters are dependent and hierarchized and to resolve the problem of the minimal action corresponding to a „world".

## 5.2. The relativistic relation between energy and momentum

In the previous sections, the expressions for the parameters energy, linear momentum and action variable for the particles of the medium and for the perturbations in the medium don't contain the values corresponding to thermal motions. If we add the thermal values, these parameters, for particles/atoms, become

$$\begin{aligned} E_a &= E_{aT} + E_{ao}, \\ \vec{p}_a &= \vec{p}_{aT} + \vec{p}_{ao}, \\ J_a &= J_{aT} + J_{ao}. \end{aligned} \tag{76}$$

Averaged in time they become:

$$\begin{aligned} \langle E_a \rangle_t &= \langle E_{aT} \rangle_t + \langle E_{ao} \rangle_t \neq 0, \\ \langle \vec{p}_a \rangle_t &= \langle \vec{p}_{aT} + \vec{p}_{ao} \rangle_t = 0, \quad \langle \vec{p}_a^2 \rangle_t = \langle \vec{p}_{aT}^2 \rangle_t + \langle \vec{p}_{ao}^2 \rangle_t \neq 0, \\ \langle J_a \rangle_t &= \langle J_{aT} \rangle_t + \langle J_{ao} \rangle_t \neq 0. \end{aligned} \tag{77}$$

For perturbations (for example, wave packet in motion), the averaged energy and action are (the medium is solid and executes thermal oscillations with maximum amplitude $q_{0T}$)

$$E_p = E_{pT} + E_{po} + E_{kp} = (\rho l_x l_y \lambda) \left[ \frac{q_{0T}^2 \omega_a^2}{2} + \frac{2\pi^2}{\alpha} \frac{q_{0p}^2}{\lambda^2} u^2 + \frac{\pi^2}{\alpha} \frac{q_{0p}^2}{\lambda^2} \upsilon^2 \right],$$

$$J_p = J_{pT} + J_{po} = (\rho l_x l_y \lambda) \frac{q_{0T}^2 \omega_a}{2} + \frac{\pi}{\alpha} \rho l_x l_y q_{0p}^2 u. \tag{78}$$

The momentum is

$$\vec{P}_p = \vec{P}_{pT} + \vec{P}_{po} + \vec{p}_{pk} = \left(\rho l_x l_y \lambda\right)\left(\dot{\vec{q}}_T + \dot{\vec{q}}_p + \frac{2\pi^2 q_{0p}^2}{\alpha \lambda^2}\vec{\upsilon}\right). \tag{79}$$

The averaged momentum in time and space is

$$\left\langle \vec{P}_p \right\rangle_{tz} = \left\langle \vec{P}_{pT} + \vec{P}_{po} + \vec{p}_{pk} \right\rangle_{tz} = \left(\rho l_x l_y \lambda\right)\left(\left\langle \dot{\vec{q}}_T \right\rangle_t + \left\langle \dot{\vec{q}}_p \right\rangle_t + \vec{\upsilon}\right) = \left(\rho l_x l_y \lambda\right)\frac{2\pi^2 q_{0p}^2}{\alpha \lambda^2}\vec{\upsilon} = m_{p\lambda}\vec{\upsilon}. \tag{80}$$

The averaged quadratic momentum is

$$\left\langle \vec{P}_p^2 \right\rangle_{tz} = \left\langle \left(\vec{P}_{pT} + \vec{P}_{po} + \vec{p}_{pk}\right)^2 \right\rangle_{tz} =$$

$$\left(\rho l_x l_y \lambda\right)^2 \left[\left\langle \dot{\vec{q}}_T^2 \right\rangle_t + \left\langle \dot{\vec{q}}_p^2 \right\rangle_{tz} + \left\langle \vec{\upsilon}^2 \right\rangle_{tz} + 2\left(\left\langle \dot{\vec{q}}_T \dot{\vec{q}}_p \right\rangle_{tz} + \left\langle \dot{\vec{q}}_T \vec{\upsilon} \right\rangle_{tz} + \left\langle \dot{\vec{q}}_p \vec{\upsilon} \right\rangle_{tz}\right)\right] = \tag{81}$$

$$\left(\rho l_x l_y \lambda\right)^2 \left[\left\langle \dot{\vec{q}}_T^2 \right\rangle_{tz} + \left\langle \dot{\vec{q}}_p^2 \right\rangle_{tz} + \vec{\upsilon}^2\right] = \left(\rho l_x l_y \lambda_0\right)^2 \frac{\vec{q}_{0T}^2 \omega_a^2}{2} + P_{p0}^2 + m_{p\lambda}^2 \vec{\upsilon}^2$$

and it is in agreement with the fact that the total energy (78) contains the thermal energy.

The observer in the "acoustic world" and his tools have no access (can't measure) the thermal part of energy (implicitly of the square of the momentum) and of action. For this observer these quantities have the expressions

$$E_{p,obs.} = E_{po} + E_{kp} = \left(\rho l_x l_y \lambda\right)\left[\frac{2\pi^2}{\alpha}\frac{q_{0p}^2}{\lambda^2}u^2 + \frac{\pi^2}{\alpha}\frac{q_{0p}^2}{\lambda^2}\upsilon^2\right],$$

$$P_{p,obs.}^2 = \left\langle \vec{P}_p^2 \right\rangle_{tz} = m_{p\lambda}^2 u^2 + m_{p\lambda}^2 \vec{\upsilon}^2 = P_{p0}^2 + p_{p\lambda}^2, \tag{82}$$

$$J_{p,obs.} = J_{po} = \frac{\pi}{\alpha}\rho l_x l_y q_{0p}^2 u.$$

In the case of a three-dimensional rest packet, $\upsilon = 0$ (the temperature of the gas made of packet is zero $T_p \to 0\mathrm{K}$), the area $l_x l_y$ becomes the same order as the square of the wavelength

$$l_x l_y \cong \lambda_0^2. \tag{83}$$

and so the energy and the action of the packet acquire the form

$$E_{p,obs.} = E_{po} \cong \left(\rho \lambda_0^3\right)\left(\frac{2\pi^2}{\alpha}\frac{q_{0p}^2}{\lambda_0^2}u^2\right),$$

$$P_{p,obs.}^2 = P_{p0}^2, \tag{84}$$

$$J_{p,obs.} = J_{po} \cong \frac{\pi}{\alpha}\rho \lambda_0^2 q_{0p}^2 u.$$

If we take into account the expressions for the quadratic momentum from the relations (83) and (84), it follows, multiplying by $u^2$,

$$P_{p,obs.}^2 u^2 = m_{p\lambda}^2 u^4 + m_{p\lambda}^2 \vec{\upsilon}^2 u^2 = E_{p0}^2 + u^2 p_{p\lambda}^2. \tag{85a}$$

or

$$E^2_{p,obs} = E^2_{p0} + u^2 p^2_{p\lambda}. \tag{85b}$$

that is a relativistic relation for energy and momentum.

### 5.3. The bound values of energy and action and their physical significance

The energy of the packet and the action cannot be however big, because the amplitude of the oscillations in the packet cannot exceed the bound of the bond breaking between the atoms of the material and the production of cavities in the fluid and melting in the solid. The limit value for the amplitude is the average length between atoms, in the fluid, or the lattice constant, in the solid, $q_{op} \leq a$. Same reasoning leads us to a bound for the wavelength which cannot be less than the double of the lattice constant $\lambda_o \geq 2a$. With these boundaries we arrive at the extreme values for energy and action

$$E_{p o \lim.} \cong \rho a^3 \frac{4\pi^2}{\alpha} u^2 = \frac{4\pi^2}{\alpha} m_a u^2 \cong E_{bd},$$

$$J_{p,\lim.} \cong \frac{\pi}{\alpha} \rho \lambda_0^2 q_{0p}^2 u = \frac{4\pi}{\alpha} a m_a u. \tag{86}$$

which are constants of the "acoustic world". In boundary conditions, we find again the relations (34 and 35) from paper [1]. It follows that the maximum energy of the packet is approximately equal to the bonding energy and the corresponding limit action. In "the acoustic world" the limit action (86) is the minimal action, because this world is made of packets and systems of packets for which the actions are added.

According to quantum theory [32, Ch. 4], the mechanical oscillations energy is quantified and the quanta is called phonon. Phonon energy is $E_{phon} = \hbar \omega$. It follows that the minimum action (86) is $\hbar$ so $J_{p,\lim.} \cong a m_a u = \hbar$. We have obtained a classical interpretation of the Planck constant [33].

## 6. Conclusions

A packet of standing mechanical waves behaves like a particle with energy localized in a volume of the average wavelength order. To this particle a rest equivalent mass connected with the rest energy through an Einstein type relation is assumed. With respect to a moving observer, the packet behaves like a particle which carries energy at the transport speed of the reference frame and in the opposite direction of its motion. The relations between the quantities of the moving packet and those of the standing packet are similar to the Lorentz transformations, the speed of light being replaced with the propagation speed of mechanical waves in the medium. The relation between the relativistic packet energy and the relativistic equivalent mass is the Einstein type. The action corresponding to the moving particle has the same expression as the standing packet action, i.e. is invariant with respect to the Lorentz transformations. The particle linear momentum, which is defined as the equivalent relativistic mass multiplied by the speed, is in a de Broglie type relation with the invariant action. Also, the relation between the particle energy and the action, both resting and in motion, are analogue to the relations that connect the same quantities in the case of an elementary particle. This relation is similar to the Planck's relation.

Studying the behaviour of the standing packet with respect to an accelerated system (Rindler transformations), we prove that the particle equivalent mass is an inertial mass (a measure of the mechanical inertia of the particle). It follows that the mechanical wave packet has the characteristics of a particle having, apparently, the parameters localized in a small volume. The behaviour of this particle, as a particle, is complete if we prove that it can be also assumed a gravitational equivalent mass that is it interacts with other mechanical perturbations which propagate in the medium in its presence. We will make this proof in a future paper.

A substantial medium, in which the mechanical perturbations propagate, behaves like an acoustical universe in which the maximum speed of energy and information propagation is the speed of the mechanical waves in an unperturbed medium. With the same speed are causally connected all the phenomena taking place in this acoustical world. The coordinates and time transformation relations are Lorentz type. The space-time far from the wave packets (stationary or in motion) is flat, the metric being minkowskian. In the neighbourhood of the wave packets, the metric is non-minkowskian, because these implyes the medium to be non-homogeneous and anisotropic. The non-homogeneity and anisotropy of the medium implyes the dependence of the waves/perturbations propagation speed on the coordinates, $u(\vec{r})$. This phenomenon is equivalent with the appearance of a refraction index of the medium dependent on the coordinates with respect to the centre of the wave packet, $n(\vec{r})$.

Interesting is the fact that the parameters of medium perturbations are proportional to the strain $\varepsilon(\vec{r},t,q_0,\lambda)$ which depends on the point, time, the perturbation maximum amplitude and the wavelength.

In the last section we also showed that the relativistic dynamical relation between energy and momentum follows from the model of the wave packet for a particle. We also showed the existence of some boundary for the energy of a perturbation and the corresponding action. In the limit of the extreme case the two quantities are the bonding energy between the components of the medium and the maximum action corresponding to a perturbation (to the wave packet) which is a constant of the „acoustical world". This maximum action in the case of the packet (a particle of the „acoustical world") becomes the minimal action in the case of the particle systems in this world.